\documentclass[prb,twocolumn,showpacs,floatfix]{revtex4-1}
\usepackage{epsfig,subfigure,amssymb,amscd,amsmath,graphicx,amsfonts}
\usepackage{times}

\newcommand{\non}{\nonumber}

\newcommand{\bea}{\begin{eqnarray}}
\newcommand{\eea}{\end{eqnarray}}
\newcommand{\be}{\begin{equation}}
\newcommand{\ee}{\end{equation}}
\newcommand{\ba}{\begin{align}}
\newcommand{\ea}{\end{align}}

\newcommand{\ZZ}{\mathbb{Z}}

\newcommand{\anc}{c^{\phantom\dagger}}
\newcommand{\crc}{c^\dagger}
\newcommand{\ana}{a^{\phantom\dagger}}
\newcommand{\cra}{a^\dagger}

\newcommand{\bk}{{\boldsymbol{k}}}

\newcommand{\bq}{{\boldsymbol{q}}}

\newcommand{\bs}[1]{ \boldsymbol{#1} }
\begin{document}

\title{Zero energy and chiral edge modes in a p-wave magnetic spin model}

\author{G. Kells$^{1}$ and J. Vala$^{1,2}$}

\affiliation{$^{1}$  Department   of  Mathematical  Physics,  National University of Ireland, Maynooth, Ireland. \\ $^{2}$ Dublin Institute for Advanced  Studies, School of Theoretical  Physics, 10 Burlington Rd, Dublin, Ireland. }

\begin{abstract}
In this work we discuss the formation of zero energy vortex and chiral edge modes in a fermionic representation of the Kitaev honeycomb model. We introduce the representation and show how the associated Jordan-Wigner procedure naturally defines the so called branch cuts that connect the topological vortex excitations. Using this notion of the branch cuts we show how to, in the non-Abelian phase of the model, describe the Majorana zero mode structure associated with vortex excitations. Furthermore we show how, by intersecting the edges between Abelian and non-Abelian domains , the branch cuts dictate the character of the chiral edge modes. In particular we will see in what situations the exact zero energy Majorana edge modes exist.  On a cylinder, and for the  particular instances where the Abelian phase of the model is the full vacuum, we have been able to exactly solve for the systems edge energy eigensolutions and derive a recursive formula that exactly describes the edge mode structure. Penetration depth is also calculated and shown to be dependent on the momentum of the edge mode.  These solutions also describe the overall character of the fully open non-Abelian domain and are excellent approximations at moderate distances from the corners. 
\end{abstract}


\date{\today} \maketitle

\section{Introduction}

Models that display spinless p-wave pairing are known to exist in both Abelian and non-Abelian topological phases. The systems are BdG (Bogoliubov de-Gennes) type topological insulators \cite{Schnyder08}, and therefore support gapless chiral modes at the edges between  Abelian and non-Abelian domains. When these edge modes have zero energy they are known to be Majorana fermions.  In addition to this the bulk of a non-Abelian phase is capable of supporting Majorana zero modes which are localized, gapped, and give rise to non-Abelian statistics \cite{ReadGreen00,Ivanov01,Stern04,Stone06, Fendley07}. 

The understanding of these properties has been greatly enhanced through the use of exactly or nearly solvable spin models. Arguably the most important for the spinless p-wave system is the Kitaev Honeycomb system \cite{Kitaev06}. The Abelian phase of model can be analyzed using perturbation theory \cite{Kitaev06,Sch07,Dus08,Vid08,Kells08} and is reduced to the so called 'Toric Code' system in this limit\cite{Kitaev03}. The main advantage of this system however is that it can be understood as either Majorana \cite{Kitaev06,Pachos07,Lahtinen08,Bas07} or Dirac fermions\cite{Feng07,Lee07,Chen07a,Chen07b,Yu08,Kells09b} hoping in a  $\ZZ_2$ gauge field. In the Dirac fermion picture, obtained using Jordan-Wigner type fermionization procedures, the spin Hamiltonian in each gauge sector reduces exactly to a mean-field type p-wave system \cite{Chen07b} but where the fermionic vacuum is exactly that of the 'Toric Code' \cite{Kells09b}.  In these paper  we will discuss  the structure of vortex and edge zero modes for the honeycomb system using this later representation.

\begin{figure}
      \includegraphics[width=.4 \textwidth,height=0.38\textwidth]{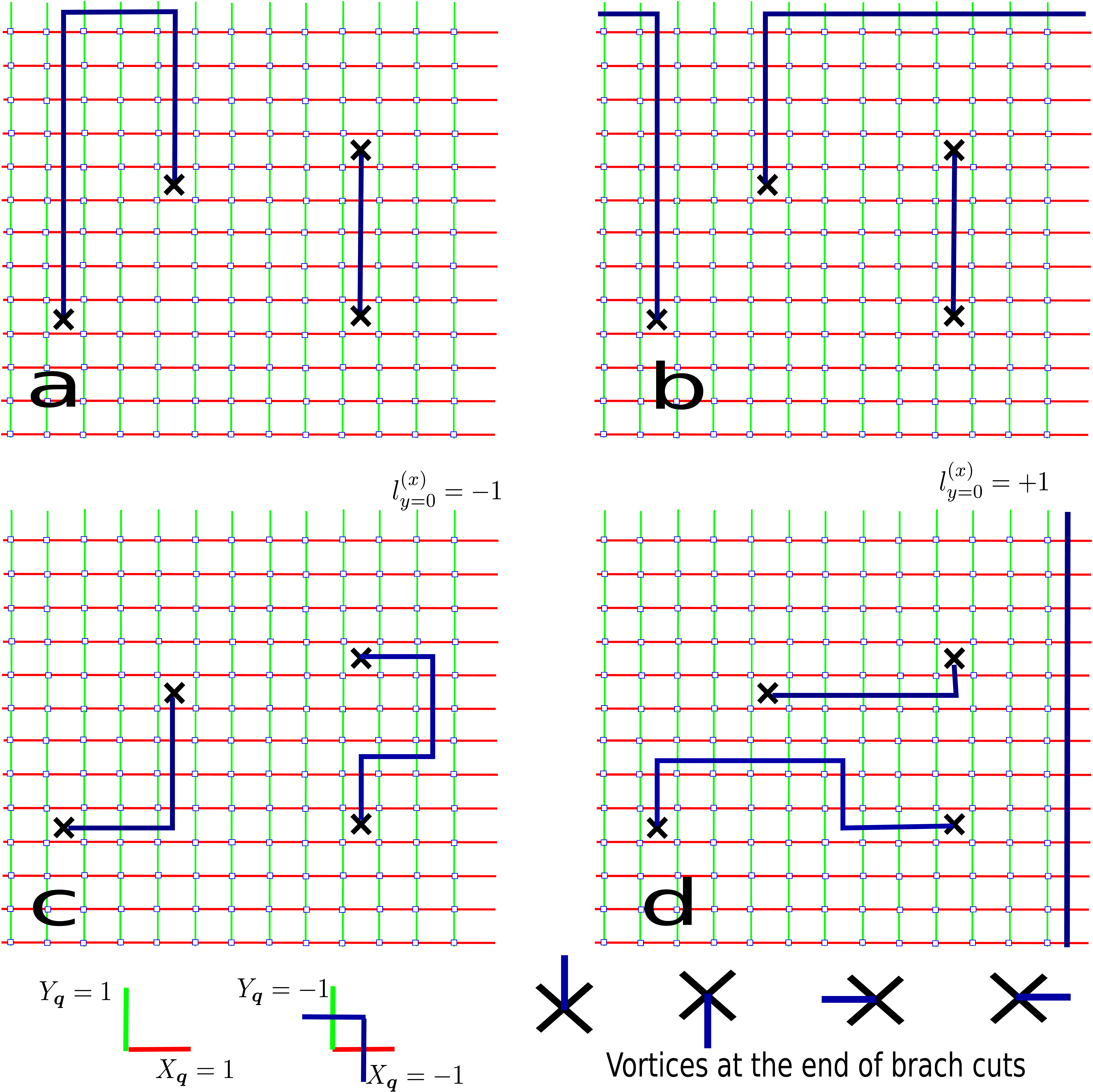}
      \caption{Vortices always appear at the end of a branch cut. Figures (a) and (b) are real vortex configurations.  Note that the eigenvalues of the homologically non-trivial symmetries dictate which vortices are connected to each other. In the absence of any vortices the $x$ and $y$ anti-periodic homological conditions are encoded as lines $X_{(N_x,y)}=-1$ $\forall y$ and  $Y_{(N_y,x)}=-1$ $\forall x$ respectively.  With the conventions used in \cite{Kells09b} the term $Y_{(N_y,0)}=-1$ dictates which vortices are connected by the branch cuts. Figures (c) and (d) are ``simulated '' vortex configurations obtained by varying couplings $J_x$ and $J_y$ and $\kappa$.}
      \label{fig:torus1}
\end{figure}
The overall aim of this paper is to present an alternative explanation for the zero and low-energy chiral modes that exist in this system. Our perspective  is complementary to previous work on the honeycomb model edge states (see for example Ref \onlinecite{Kitaev06} Appendix B and Ref. \onlinecite{Lee07}), the continuum p-wave analysis of Refs. \onlinecite{ReadGreen00,Fendley07}, and the bosonic condensation theory presented in Refs. \onlinecite{Bais08a, Bais08b}.

The first half of the paper describes how the notion of a branch cuts arise naturally from the 2-D Jordan-Wigner procedure. This is in contrast with mean field p-wave analysis where the branch cuts are an afterthought to ensure that the modes are single-valued.  We will see that, as expected, these branch cuts connect the topological defects (vortices) of the system. However, through out this story we will attempt to emphasize that it is the branch cuts that are the fundamental objects. For example, it is the branch cuts, and not the vortices, that dictate the fermionic behavior of the system. This perspective also holds on the boundaries between Abelian and non-abelian domains. For example we will see that it is the number of branch cuts through those edges that dictate the character of the modes found there.  In the second half of the paper we will analyse the zero energy bulk modes and the zero energy and chiral edge modes found in the model.  Our analysis of edge modes is valid for both cylindrical and fully open boundary conditions but is based on the consistency relations between homologically trivial excitations (vortices) and the  homologically non-trivial excitations on a torus \cite{Kells09b}. We first introduce the cylindrical system and describe the general character of the modes found in this case. The general conclusion is that exact zero modes only form on edges that are intersected by an even number of branch cuts. In addition to this we see for the hard boundary condition (i.e. where the Abelian domain is exactly the vacuum), that there are exact solutions for the BdG equations. We use these solutions to examine the mode penetration depth as a function of the Hamiltonian parameters and the mode momenta along the edge.

We finally extend the general analysis to fully open rectangular boundary conditions  we see that  exact zero modes only form in this case when there is an {\em odd} number of branch cuts through the domain. The reason for the difference is an extra phase factor that is contributed at the corners of the system. At moderate distances from the corners however the exact solutions for the cylindrical hard boundary system are an excellent approximation for the open system eigenmodes.  These results are general agreement with Ref. \onlinecite{Kitaev06} and Ref. \onlinecite{ReadGreen00}.

\section{Fermionic Formulation}
It was shown in \cite{Kells09b} that each vortex sector of the honeycomb lattice model can be written as 
\be
H=H_0+\sum_\bq \sum_l P_\bq^{(l)} \non
\ee
where in terms of fermions we can write
\bea 
H_0 &=& J_x \sum_{\bq} \bs{X}_{\bq} (\crc_{\bq}-\anc_{\bq})
(\crc_{\bq\rightarrow}+\anc_{\bq\rightarrow}) \non \\ 
 &+& J_y \sum_{\bq} \bs{Y}_\bq (\crc_{\bq}-\anc_{\bq} ) (
\crc_{\bq\uparrow}+\anc_{\bq\uparrow} )\non \\ 
 &+& J_z \sum_{\bq} (2\crc_{\bq} \anc_{\bq} - I) ,
\label{eq:Hc}
\eea
where we have used the shorthand $\bq\rightarrow=\bq+n_x$, $\bq \uparrow = \bq+n_y$ and $\bq \nearrow = \bq+n_y+n_x$. In the plane, $\bs{Y}_\bq= I $ for all $\bq$ and $\bs{X}_\bq$ is defined as
\be
\label{eq:X}
\bs{X}_{x,y} \equiv \prod_{y'=0}^{y-1} \bs{W}_{x,y'} . 
\ee

The terms $P^{(l)}$ are explicit $T$-symmetry breaking terms, the fermionic form of which was also derived in $\cite{Kells09b}$. For simplicity in this work we will retain only terms $P^{(1)},P^{(2)}, P^{(3)}$ and $P^{(4)}$.  These terms are sufficient to generate the required non-abelian phase and lead to more symmetrical solutions. Explicitly these terms are \cite{Kells09b}:
\bea
P^{(1)}&=& -i \kappa  \bs{X}_{\bq} (\crc_{\bq}-\anc_{\bq})(\crc_{\bq\rightarrow}-\anc_{\bq\rightarrow}) \\
P^{(2)}&=& -i \kappa  \bs{X}_{\bq\uparrow} (\crc_{\bq\uparrow}+\anc_{\bq\uparrow})(\crc_{\bq\nearrow}+\anc_{\bq\nearrow}) \\
P^{(3)}&=&  + i \kappa  \bs{Y}_{\bq} (\crc_{\bq}-\anc_{\bq})(\crc_{\bq\uparrow}-\anc_{\bq\uparrow}) \\
P^{(4)}&=&  + i \kappa  \bs{Y}_{\bq\rightarrow} (\crc_{\bq\rightarrow}+\anc_{\bq\rightarrow})(\crc_{\bq\nearrow}+\anc_{\bq\nearrow})
\label{eq:HP}
\eea

The Jordan-Wigner convention used to define the fermions is directly responsible for how vorticity is encoded in the fermionic system. For the string convention chosen in \cite{Kells09b} the vorticity is encoded in the fermionic Hamiltonian through the condition (\ref{eq:X}). On a torus there are additional homologically non-trivial degrees of freedom which also need to be determined consistently with the condition (\ref{eq:X}). These homologically non-trivial are encoded the $X_\bq$ and $Y_\bq$ values at the boundary of the system \cite{Kells09b}.  Recently we have extended this Jordan-Wigner method to deal with the Yao-Kivelson 3-12 lattice variant of the model \cite{Kells10}.

The consistency relations provided in Ref. \onlinecite{Kells09b} have an interesting pictorial representation which leads us naturally to the concept of branch cuts and a less restrictive understanding of vorticity.  For any vortex arrangement we see that there are lines of $X_\bq=-1$ and $Y_\bq=-1$ which together connect vortices in pairs. In Figure \ref{fig:torus1} we have provided a number of examples. 

On an open plane we no longer have these homologically non-trivial symmetries but neither do we have the condition that vortices are created in pairs: $\prod_\bq W_\bq =1$. In this case valid vortex sectors can be encoded using the following guidelines. 
\begin{itemize}
 \item The vortex free sector ($W_\bq=1 \forall \bq$) is encoded as $X_\bq=1 \forall \bq$. 
 \item A single isolated vortex at position $\bq$ is encoded with $X_\bq =1$ everywhere except for a single line of $X_{x,y}=-1$ starting at $y+n_y$ and extending to infinity.
 \item When two vortices occur at different $x$-positions there are two unique strands of $X_\bq=-1$ connecting them both to infinity.
 \item If two vortices occur at different $y$-positions but with the same $x$ a line of $X_\bq=-1$ connects them together.
\end{itemize}

One can `simulate' the change of vortex sectors by altering the coupling constants (the $J_x$ and $J_y$) on unique links \cite{Lahtinen09}. Thus by changing the sign of $J_x$ at $q$ one effectively changes the gauge encoding  $X_\bq$. Strictly speaking this does not change the vortex sector of the Hamiltonian however. With our fermionization convention, and on a plane, there is no vortex sector which would correspond to the change $J_y \rightarrow -J_y $ at $\bq$. 

From now on we will take $J=J_x=J_y$, dropping the subscript and take the viewpoint used in \cite{Lahtinen09} where, by changing the coupling strengths, we can simulate changing the vortex configurations. In what follows however, and only for convenience, we will generally continue to regard the $J$  and $\kappa$ terms as constant across the lattice and allow vorticity to be encoded in the $X$ and $Y$ terms. With this perspective it is easier to appreciate that truly meaningful objects in this story are not the vortices themselves but the connected strings of $-1$'s defined on the $X_\bq$ and $Y_\bq$ matrices. Indeed as we have already shown these strings take on the role of branch cuts in our fermionic Hamiltonian and will see later that it is their ends that give rise to localized zero modes.  From this perspective we can say that zero modes are only associated with vortices because a branch cut always happens to end there. 

In addition to the vortex zero-modes we will also see in what follows that it is the branch cuts that are directly responsible for the appearance of the single extended zero mode that occurs at the interface between abelian and non-abelian phases when an odd number of (ordinary localized) zero-modes are in the non-abelian bulk. The parameter $J$ dictates which phase we are in. For $J<J_z/2$ we are in the abelian phase and for $J>J_z/2$ we are in the non-Abelian phase if $\kappa \ne 0$. In what follows we will specify the $J$ and $\kappa$ values in the Abelian domains as $J_A$ and $\kappa_A$ respectively. 

\section{Bulk Majorana fermion zero modes}
In this section we will briefly discuss the bulk Majorana modes found at the end of the branch-cuts. We will not however discuss the detailed structure of the bulk modes other than to present some numerical calucations. In later sections however we will demonstrate how the structure can be seen as a limiting case of edge modes found between domains of Abelian and non-Abelian topological phase.  

We begin by presenting the Bogoliubov-De Gennes formalism. The full position space Hamiltonian can be written in the form
\bea
H=  \frac{1}{2} \sum_{\bq \bq'} \left[\begin{array}{cc} c^\dagger_{\bq} & c_\bq
\end{array}  \right] \left[
\begin{array}{cc} \xi_{\bq \bq'} & \Delta_{\bq \bq'} \\ \Delta^\dagger_{\bq
\bq'} & -\xi^{T}_{\bq \bq'} \end{array} \right] \left[\begin{array}{c}
c_{\bq'} \\ c^\dagger_{\bq'} \end{array} . \right] 
\label{eq:Hg} 
\eea

This system can be diagonalized by solving the Bogoliubov-De Gennes eigenvalue problem
\bea
\label{bdgtransf}
\left[ \begin{array}{cc} \xi & \Delta \\ \Delta^\dagger & -\xi^{T} \end{array} \right] =  \left[\begin{array}{cc} U & V^* \\ V & U^* \end{array}  \right]  \left[ \begin{array}{cc} E & \bs{0} \\ \bs{0} & -E \end{array}  \right] \left[ \begin{array}{cc} U & V^* \\ V & U^* \end{array}   \right]^\dagger,
\label{eq:BdG}
\eea
where the non-zero entries of the diagonal matrix $E_{nm} = E_n \delta_{nm} $ are the the quasi-particle excitation energies. The Bogoliubov-Valatin quasi-particle excitations are 
\bea
\label{eq:gamma}
&&\left[\begin{array}{cc}a_1^\dagger,...,a_M^\dagger,  & a_1,...,a_M \end{array}  \right] \\&& =  \left[\begin{array}{cc} c_1^\dagger,..., c_M^\dagger, & c_1,...,c_M \end{array}  \right]  \left[\begin{array}{cc} U & V^* \\ V & U^* \end{array}  \right]  .
\eea
which after inversion and substitution into (\ref{eq:Hg}) give
\be
H= \sum_{\bs{n}=1}^{M} E_n (a^\dagger_n a^{\phantom \dagger}_n - \frac{1}{2}) .
\label{eq:Hd}
\ee

In spinless p-wave systems  it is guarunteed by an index theorem that in the the case of the $2N$ well separated vortices we have $2N$ zero energy $(E=0)$ fermionic modes of which $N$ must be identified as $a^\dagger$'s and $N$ as $a$'s\cite{Volovik93}.  It is rather remarkable that one can always choose a superposition of the $2N$ $a^\dagger$ and $a$ zero-modes such that the resulting modes are fully localized around the vortex excitations.  
\bea 
\gamma_j &=& \sum_{n=1}^{N} \alpha_{jn} \cra_n + \alpha_{j,n+N} \ana_n \\ &=& \left[\begin{array}{cc} c_1^\dagger,..., c_M^\dagger, & c_1,...,c_M \end{array}  \right] \left[\begin{array}{c} u_{\bq,j} \\ v_{\bq,j}  \end{array} \right] .                                                                                                                                                                                                                                                                                                                     \eea
\begin{figure*}[ht]
\centering
\subfigure[]{
\includegraphics[width=.45\textwidth,height=0.24\textwidth]{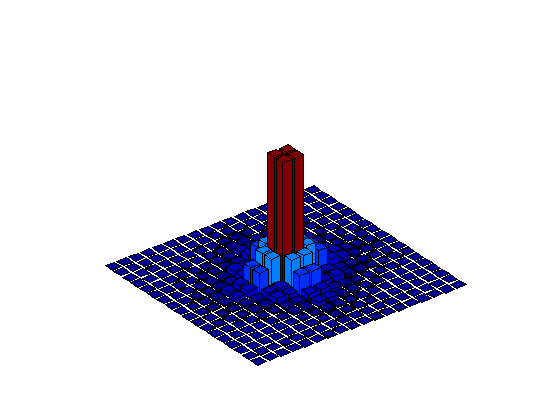}
\label{fig:vs1}
}
\subfigure[]{
\includegraphics[width=.45\textwidth,height=0.24\textwidth]{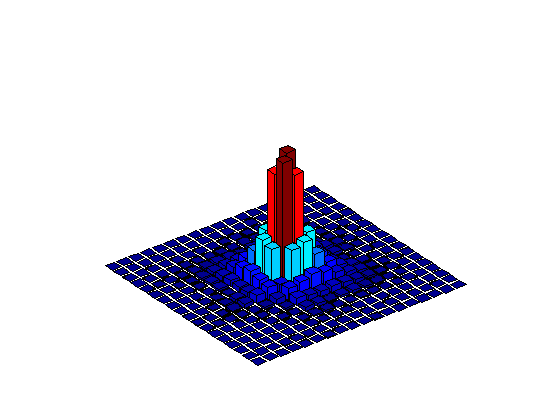}
\label{fig:vs2}
}
\caption[]{(Color online) The position space structure $|u_\bq|=|v_\bq|$ of vortex Majorana zero-modes for (a) $J_z=1$, $J=1$ and $\kappa=0.5$ and (b) $J_z=1$ $J=0.8$ and $\kappa=0.2$}
\label{fig:vstructure}
\end{figure*}
It is interesting to note that this localization condition also enforces the condition that $u_{\bq j}= e^{i \Omega_n} v_{\bq j}^*$. However, if one wishes to call this a Majorana mode $\gamma_j=\gamma_j^\dagger$ it is necessary to multiply the states $(u,v)^T$ by the overall phase $e^{-i \Omega_n/2}$  such that $u_{\bq,j} = v_{\bq,j}^*$. It was pointed out by Stone and Chung \cite{Stone06} that the Majorana condition therefore fixes the global phase of the states $\gamma_j$ up to an overall sign $\pm 1$. Understanding how and when this overall sign changes is crucial to understanding how non-abelian statistics arise in this degenerate subspace. In Figure \ref{fig:vstructure} we show the $|u_\bq|$ and $|v_\bq|$ position space structure for some different values of $J$ and $\kappa$.  

\section{Unpaired Majorana modes and edge states}

The non-Abelian phase of Kitaev models are Topological insulators of the BdG class, see for example \cite{Schnyder08}. Roughly speaking this means that we have a bulk energy spectrum which is `insulating' (does not cross the Fermi energy at $E=0$) and an edge spectrum which is `conducting' (does cross the Fermi energy at $E=0$).  For a careful choice of edge conditions it is possible to analytically treat the `conducting' edge modes. 

In order to determine the structure of the modes let us consider an element of an arbitrary eigenstate $a^\dagger_n$ of the BdG Hamiltonian.  Each value  $u_{x,y}$ is connected to  $u_{x\pm 1, y}$ and $u_{x, y \pm 1}$ through the non-zero elements of the $\xi$ matrix and to $v_{x \pm 1, y }$ and $v_{x , y \pm 1}$   through the non-zero elements of the $\Delta$ matrix. It is quite difficult to say anything generic about the form that an eigenvector should have. One feature is universal however. We see that if the elements around the point $u_{x,y}$ are almost zero then $u_{x,y}$ should also be almost zero. It is true regardless of the values we give our coefficients in our Hamiltonian and it is this rule that determines the vast majority of the zero-mode structure (or lack of it).

In the absence of branch cuts, there is a simple condition that the nine interconnected elements must obey if they are to be eigenstates of the system:
\bea
&& (2 J_z -E) u_{x,y} + \\ \non && J( u_{x+1,y} +u_{x-1,y}+ u_{x,y+1} + u_{x,y-1}) +\\ \non && 
   (J- 2 i \kappa) v_{x+1,y}  + (-J + 2 i \kappa ) v_{x-1,y} + \\ \non && (J + 2 i  \kappa ) v_{x,y+1} + (-J - 2 i  \kappa ) v_{x,y-1} = 0.
\label{eq:condition}
\eea

For edge states on a cylinder we make the reasonable assumption is that, in the direction of edge, our modes are plane waves (momentum eigenstates). For example along the lower edge of  a cylindrical non-Abelian domain we have BdG excitations of the form 
\be
a_n^\dagger = \mathcal{N} \sum_\bq  e^{\pm i k_x x} (u(y-y_0) \crc_{\bq} +  v(y-y_0) \anc_{\bq})
\ee
This state corresponds to a superposition of left (right) moving particles and right (left) moving holes. On a cylinder the allowed values of $k_x$ are $ 2 n \pi /N_x$ when there is an even number of branch cuts through the edge and $ 2 (n+1/2) \pi /N_x$ when the number is odd. The basic reasoning is this.  A branch cut is accommodated in (\ref{eq:condition})  by a change in signs of the elements $J$ and $\kappa$ acting on some (not all) of the values $u$ and $v$.  To keep the energy low then the phase of the mode $a_n^\dagger$ should abruptly change sign to counteract the sudden sign change in the fermionic Hamiltonian.  
 
\begin{figure*}[ht]
\centering
\subfigure[]{
\includegraphics[width=.3\textwidth,height=0.2\textwidth]{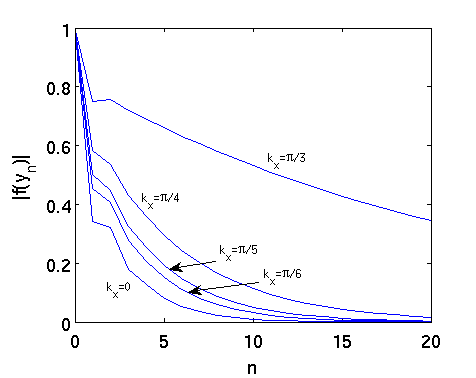}
\label{fig:absdepth}
}
\subfigure[]{
\includegraphics[width=.3\textwidth,height=0.2\textwidth]{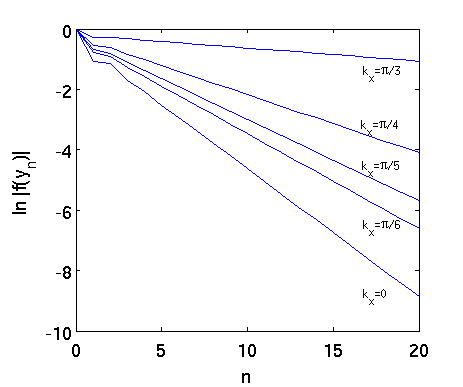}
\label{fig:logdepth}
}
\subfigure[]{
\includegraphics[width=.3\textwidth,height=0.21\textwidth]{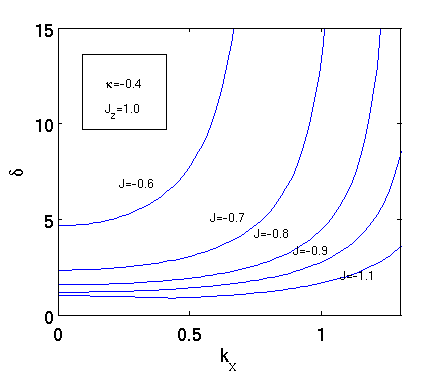}
\label{fig:depthJ}
}
\caption[]{(Color online) (a) The function $|f(y_n)|$ for different  $k_x$ with $J_z=1, J=-.7$ and $\kappa=-.4$. (b)  A log plot of the same function $f(y_n)$ again with $J_z=1, J=-.7$ and $\kappa=-.4$. (c) The penetration depth $\delta$ as a function of $k_x$ for different values of $J$ and fixed $J_z$ and $\kappa$. Penetration depth goes to infinity approximately when $|d_1+d_2|> 1$ }
\label{fig:depth}
\end{figure*}

On a cylinder this has interesting consequences. Let us start from the toroidal case and open up the $y$-boundary above and below the $y=0$ line. We now have two edges which are some distance apart. Translation invariance remains in the $x$-direction but is broken in the $y$-direction. Recall now that the anti-periodic $x$-boundary condition is encoded as a single line of $X_\bq=-1$. Thus in the periodic vortex free sector we therefore have chiral edge states with $k_x =\pm  2 n \pi /N_x$. This includes {\em two} edge zero-modes, one on each edge. In the anti-periodic vortex free case we have no zero modes.  This is because we have a single branch cut intersecting both edges and thus $k_x= \pm  2 (n+1/2) \pi /N_x$.

If a single vortex exists inside the cylinder there must be a branch cut connecting it to either infinity or some other vortex outside the cylinder. If we were originally in the periodic system then the introduction of a branch cut through one wall would destroy periodicity on this edge and we could not have Majorana zero modes. The other edge however would remain unaffected. In the opposite sense if we were originally in the anti-periodic sector then the introduction of a vortex would restore periodicity to one of the edges and thus allow values of $k_x =\pm  2 n \pi /N_x$ to propagate along this wall. 

We can extend this reasoning to deal with fully open boundaries (non-Abelian domains within Abelian domains and vice versa). However it is useful to first solve the system exactly on a hard interface $J_A=0$ where the Abelian side of the edge is the full vacuum.  In this scenario numerical calculation shows that all low-energy modes satisfy  $u_{\bq} = e^{i \theta} v_\bq$. Thus for modes along the lower edge at $y=y_0$ we have

\be
a_n^\dagger = \mathcal{N} \sum_\bq f(y-y_0) e^{\pm i k_x x} (e^{-i\theta/2} \crc_{\bq} +  e^{+i \theta/2} \anc_{\bq})
\label{eq:ansatz}
\ee

Note that under the conditions $k_x=0$ and $Im(f)=0$ this ansatz is already a Majorana fermion.  If one now substitutes this expression into (\ref{eq:condition}) we observe that  
\be
\label{eq:Ekx}
E(J, \kappa, k_x) = \frac{8 J \kappa}{\sqrt{J^2 +4 \kappa ^2}} \sin k_x ,
\ee
and that, along the bottom edge, $\theta = \tan^{-1} (2 \kappa/J) $. Furthermore one sees that the function $f$ follows from the recursive relation 
\be
f(y_{n+2})= \frac{1}{\sqrt{J^2+4 \kappa^2}-J} [d_1 f(y_{n+1}) +d_2 f(y_n)]
\label{eq:recursive}
\ee
where 
\bea
\non
d_1 &=& 2 J_z + 2 J \cos(k_x) - i 2 \frac{J^2 -4\kappa^2}{\sqrt{J^2+4\kappa^2}} \sin(k_x)  \\ \non
d_2 &=& \sqrt{J^2+4 \kappa^2 }+J
\eea

Interestingly the structure of the mode depends on the parameter  $J_z$  but the associated energy does not.  However this feature is  present for the ($J_A = 0$) hard boundary condition only. Indeed numerical calculation shows that even the $\sin(k_x)$ dependence is not exact once the hard boundary condition is relaxed ($J_A \ne 0$).
\begin{figure}
      \includegraphics[width=.3 \textwidth,height=0.43\textwidth]{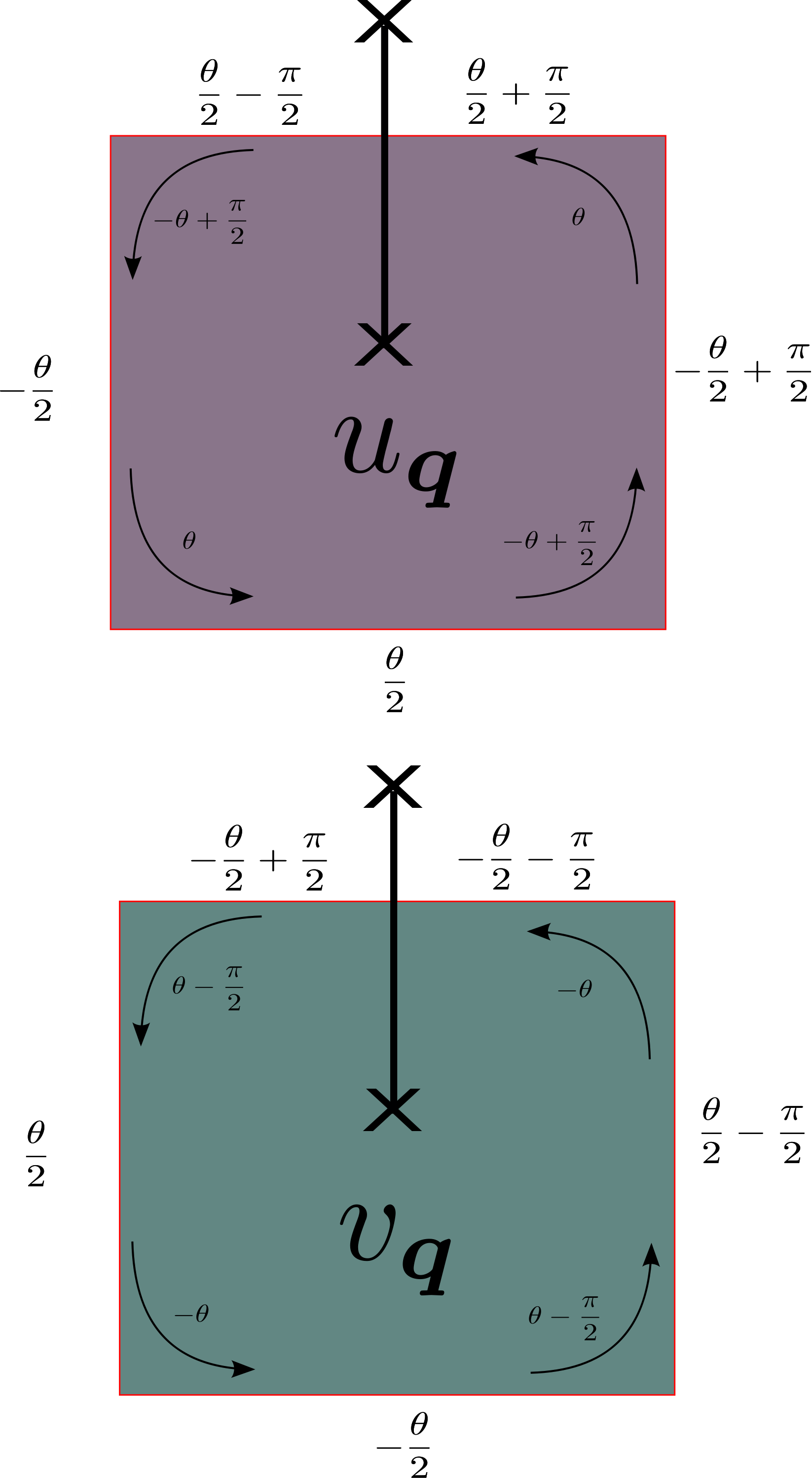}
      \caption{A schematic of how $\theta$ in the Majorana edge zero mode varies around an isolated domain of non-Abelian phase. In this model $\theta = \tan^{-1}(2 \kappa/J)$. Inside the bulk, and at the corners, we indicate the phase that must be picked up as we move around that corner in the direction indicated by the arrows. }
      \label{fig:PhaseZero}
\end{figure}

The mode penetration depth can be calculated easily from the recursive relationship (\ref{eq:recursive}), see for example FIG. \ref{fig:depth}. The most salient point is that this depth depends on $k_x$ and therefore on $E$. Loosely speaking we can say that the further the energy is from $E=0$ the further it extends into the bulk. An upper limit for the momenta $k_x$ of the edge modes can be calculated from the condition that $|d_1+d_2|< 1$. Note that this condition also says that we must be inside the non-Abelian domain $|J|>|J_z|/2$ for the solution to be normalized. 
\section{Fully open boundary conditions}
\begin{figure*}[ht]
\centering
\subfigure[]{
\includegraphics[width=.3\textwidth,height=0.2\textwidth]{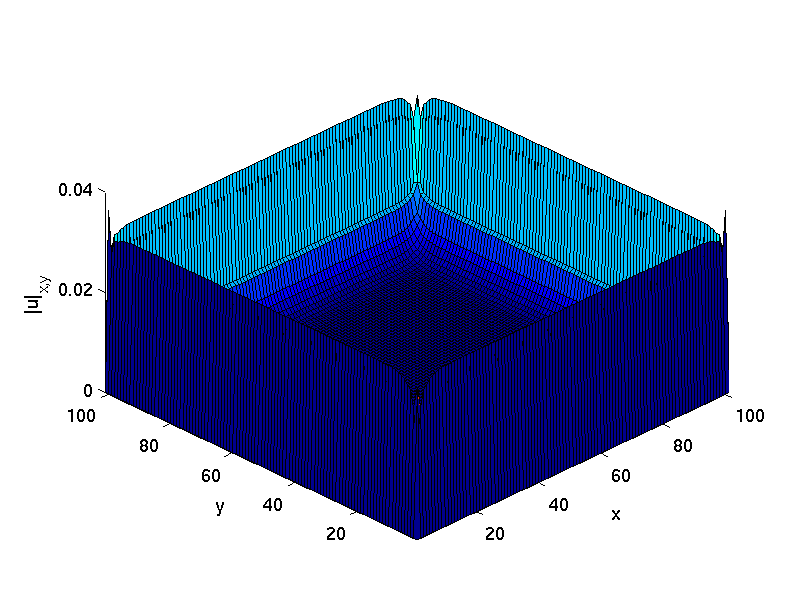}
\label{fig:OZA}
}
\subfigure[]{
\includegraphics[width=.3\textwidth,height=0.2\textwidth]{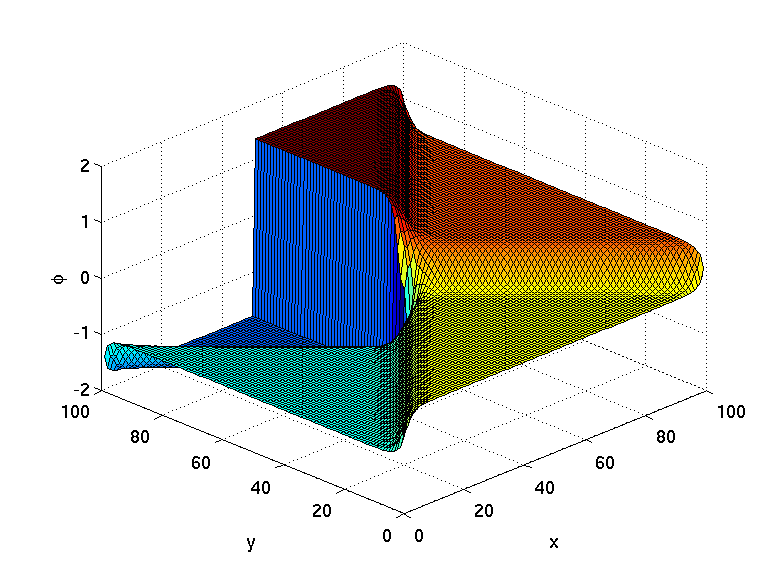}
\label{fig:OZP}
}
\subfigure[]{
\includegraphics[width=.3\textwidth,height=0.2\textwidth]{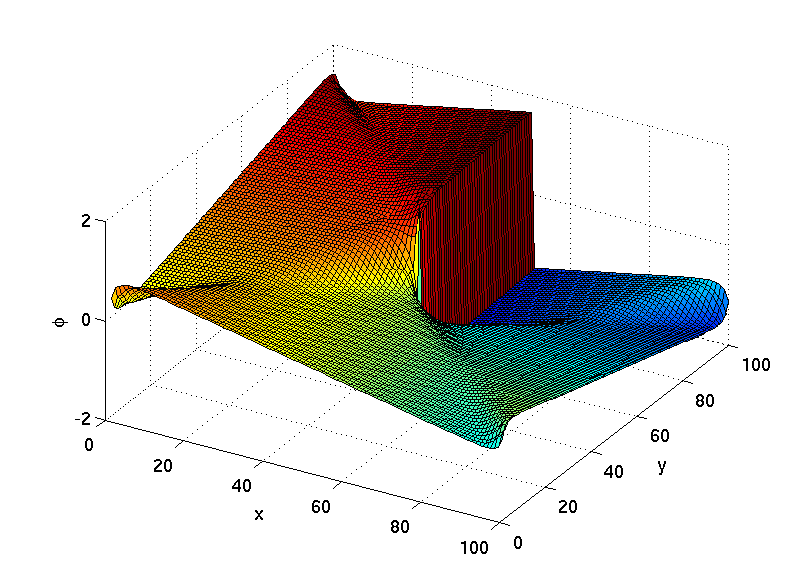}
\label{fig:OKP}
}
\caption[]{(Color online) (a) The position space structure $|u_\bq|$ of the Majorana edge zero-mode with a single vortex in the bulk (b) The phase dependence $\phi=\theta$ for the same Majorana zero mode  (c) The phase dependence $\phi=\theta+\bk \cdot \bq$ for a chiral edge mode where a  vortex exists in the bulk. We can see here the combined effects of the branch cut, the phase jumps at corners and the almost constant momenta $|k|\approx 2 \pi /L_{Total}$. In these figures $J_x=J_y=J=-0.7, J_z=1, \kappa=-.4$}
\label{fig:open1}
\end{figure*}

If we surround a non-Abelian domain  with an Abelian domain we have no zero energy states if there are no vortices inside the non-Abelian domain. If we place an odd number of vortices inside the non-Abelian domain then we do have one zero energy edge mode even though an odd number of branch cuts intersect the domain wall. 

The key to understanding all this is that phases are also picked up when the wall direction is changed  and that these phases all add up to $\pi$, canceling the branch cut phase.   A schematic of the phases picked up for the zero mode in a rectangular shaped system is shown in Figure \ref{fig:PhaseZero}.  This picture can be arrived at by analyzing each of the edges separately and assuming the appropriate plane wave momentum eigenstate (\ref{eq:ansatz}) along each edge. The trigonometric identity $\tan^-1 a/b +\tan^-1 b/a = \pi/2$ is the key to understanding why the total phase due to the corners is $\pi$. At this time we have been unable to fully resolve the exact behavior at the corners. However the numerically calculated example provided in  Figure \ref{fig:PhaseZero} shows that the phase changes at a corner happen abruptly and that the momentum eigenstates structure \ref{eq:ansatz} is rapidly returned to as we move away from the corner.  

The chiral (non-zero energy) edge modes are also similar to that seen on the cylinder.  In these cases, as for the zero momentum/energy modes, abrupt phase shifts are seen at the corners although in this case we cannot separate phase shifts due to momenta and those due to the corners. However it is worthwhile to note that if we use the value of momenta measured far from the corner in expression (\ref{eq:Ekx}) we obtain the numerically calculated energy eigenvalue for the mode exactly. This measured value of momenta is however not exactly $2 \pi n/ L_{Total}$ but slightly different magnitude. One could think of this as arising because the chiral mode sees a slightly different perimeter $L_{Total}-\Delta_L$ but we advise against taking this too literally.

The picture above can be immediately applied to domains of Abelian phase inside a non-Abelian one. If there is no vortex inside this abelian domain then there is no branch cut and all modes are chiral but where the direction of the momenta for positive and negative energy modes is in the opposite sense to that on the outer edge. If an odd number of vortices exist inside the internal Abelian domain then we have an odd number of branch cuts and a zero mode can exist. As suggested by Read and Green \cite{ReadGreen00} the zero mode due to a single vortex in the non-Abelian domain can be viewed as special case of this scenario where the domain edge has been reduced to a single plaquette. 

\section{Conclusion}
We have analysed edge mode structure of the Kitaev Honeycomb model using a Jordan-Wigner fermionization procedure. We see that the branch cuts are  naturally defined for us with the single particle Hamiltonian $\xi$ and the order parameter $\Delta$. We then extended the notion of these branch cuts to account for edge effects between Abelian and non-Abelian domains. Although our general conclusions are in agreement with other methodolgies we feel there is an inherent simplicity to the above arguements that make them an important part of the overall story. 

For the specific model we have chosen we have been able to derive a simple recursive relation that exactly dictates the structure on edge between a vacuum and non-Abelian domain. A number of key features are present. Firstly the solutions are only normalized in the non-Abelian domain. Secondly we see a clear dependence on penetration depth on the mode momenta. We have also outlined how to apply the cylindrical solutions for the hard boundary to a fully open system. 

In future work we will attempt to analyse the edge mode momentum dependency further and to extend these results to softer boundaries. We will also attempt to identify enough properties to exactly formulate the mode structure at the corners. 

\section{Acknowlegments}
We thank Steve Simon, Joost Slingerland and Ivan Rodriguez for interesting discussions.  This work has been supported by Science Foundation Ireland through the President of Ireland Young Researcher Award 05/YI2/I680.


\begin{thebibliography}{15}
\bibitem{Signs} We will usually choose the convention that $J_z = 1$, and $J_x, J_y, \kappa \le 0$. This convention has the advantage that our basic $c$ fermions are associated with anti-ferromagnetic configurations of the $z$-dimers and our vacua are toric code states on an effective square lattice \cite{Kells09b}. It allows us to naturally interpret the $\xi$ part of the Bogoliubov De-Gennes Hamiltonian as the single particle equation. In the momentum space for example this gives $E_\bk \approx c + d (k_x^2+ k_y^2)$
\end{thebibliography}
\end{document}